\documentstyle[11pt]{article}

\font\fr=eufm10 scaled \magstep 1 
\font\es=msbm10                   
\newtheorem{teor}{Theorem}
\newtheorem{prop}{Proposition}
\newtheorem{definition}{Definition}
\newtheorem{lem}{Lemma}
\def\beq{\begin{equation}}
\def\eeq{\end{equation}}
\def\bea{\begin{eqnarray}}
\def\eea{\end{eqnarray}}
\def\beann{\begin{eqnarray*}}
\def\eeann{\end{eqnarray*}}
\def\ben{\begin{enumerate}}
\def\een{\end{enumerate}}
\def\bit{\begin{itemize}}
\def\eit{\end{itemize}}
\def\dst{\(\displaystyle}
\def\derpar#1#2{\frac{\partial{#1}}{\partial{#2}}}

\def\coor#1#2#3{{#1}^{#2}, \ldots, {#1}^{#3}}
\def\moment#1#2#3{{#1}_{#2}, \ldots, {#1}_{#3}}

\def\qed{\ifvmode\removelastskip\fi
{\unskip\nobreak\hfil\penalty50\hbox{}\nobreak\hfil
\hbox{\vrule height1.2ex width1.2ex}\parfillskip=0pt
\finalhyphendemerits=0 \par\smallskip}}
\def\vf{\mbox{\fr X}}
\def\df{{\mit\Omega}}
\def\Lag{{\cal L}}

\def\d{{\rm d}}

\def\Real{\mbox{\es R}}

\def\inn{\mathop{i}\nolimits}
\def\Tan{{\rm T}}

\def\ls{((E,M;\pi),\Lag )}
\def\lag{\pounds}
\def\Cinfty{{\rm C}^\infty}
\def\proof{( {\sl Proof} )\quad}

\def\tabaddress#1{{\small\it\begin{tabular}[t]{c}#1
\\[1.2ex]\end{tabular}}}

\title{ON THE MULTIMOMENTUM BUNDLES
 AND THE LEGENDRE MAPS IN FIELD THEORIES}
\author{\sc A. Echeverr\'{\i}a-Enr\'\i quez,
M.C. Mu\~noz-Lecanda
\thanks{{\bf e}-{\it mail}: MATMCML@MAT.UPC.ES},
N. Rom\'an-Roy
\thanks{{\bf e}-{\it mail}: MATNRR@MAT.UPC.ES}
   \\
  \tabaddress{\it Departamento de Matem\'atica Aplicada y Telem\'atica.
 Edificio C-3, Campus Norte UPC.\\
   C/ Jordi Girona 1.
   E-08034 Barcelona. SPAIN}}
\date{To be published in {\sl Reports on Mathematical Physics}\\
       (math-ph/9904007)}

\begin{document}
\maketitle
\thispagestyle{empty}
\setcounter{page}{0}

\begin{abstract}
We study the geometrical background of the Hamiltonian formalism
of first-order Classical Field Theories. In particular, different
proposals of {\sl multimomentum bundles} existing in the usual
literature (including their canonical structures) are analyzed and
compared. The corresponding {\sl Legendre maps}   are introduced.
As a consequence, the definition of {\sl regular} and {\sl
almost-regular} Lagrangian systems is reviewed and extended from
different but equivalent ways.
\end{abstract}

\bigskip\bigskip\bigskip

{\bf Key words}: Jet Bundles, Classical Field Theories, Legendre
map, Hamiltonian formalism.

\bigskip

\vbox{AMS s.\,c.\,(1991): 53C80, 55R10, 58A20,70G50, 70H99.}\null

\clearpage

\section{Introduction}

The standard geometric structures underlying the covariant
Lagrangian description of first-order Field Theories are first
order jet bundles $J^1E\stackrel{\pi^1}{\to}E\stackrel{\pi}{\to}M$
and their canonical structures \cite{EMR-96}, \cite{Gc-73},
\cite{GMS-97}, \cite{Kr-87}, \cite{Kp-80}, \cite{LR-88},
\cite{Sa-89}. Nevertheless, for the covariant Hamiltonian
formalism of these theories there are several choices for the
phase space where this formalism takes place. Among all of them,
only the  {\sl multisymplectic} models will deserve our attention
in this work. So, in \cite{Go-91b}, \cite{GIMMSY-mm}, \cite{KT-79}
and \cite{MS-97}, (see also \cite{Ki-73}, \cite{KS-75} and
\cite{KT-79}) the {\sl multimomentum phase space} is taken to be
${\cal M}\pi\equiv\Lambda_1^m\Tan^*E$, the bundle of $m$-forms on
$E$ ($m$ being the dimension of $M$) vanishing by the action of
two $\pi$-vertical vector fields. In \cite{CCI-91}, \cite{Ka-97a},
\cite{LMM-95} and \cite{LMM-96} use is made of
$J^1\pi^*\equiv\Lambda_1^m\Tan^*E/\Lambda_0^m\Tan^*E$ as the
multimomentum phase space (where $\Lambda_0^m\Tan^*E$ is the
bundle of $\pi$-semibasic $m$-forms in $E$). Finally, in
\cite{EM-92}, \cite{GMS-95}, \cite{GMS-97}, \cite{Sd-95}, and
\cite{SZ-93} the basic choice is the bundle
 $\Pi\equiv\pi^*\Tan M\otimes{\rm V}^*(\pi)\otimes\pi^*\Lambda^m\Tan^*
M$
 (here ${\rm V}^*(\pi)$ denotes the dual bundle of  ${\rm V}(\pi)$:
 the $\pi$-vertical subbundle of $\Tan E$)
 which, in turns, is related to
 $J^1E^*\equiv\pi^*\Tan M\otimes\Tan^*E\otimes\pi^*\Lambda^m\Tan^* M$.
 The origin of all these multimomentum bundles is related to the
 different {\sl Legendre maps} which arise essentially from the
 fiber derivative of the Lagrangian density (see Section
 \ref{lm}).

All these choices have special features. So, ${\cal M}\pi$ and
$J^1E^*$ are endowed with natural multisymplectic forms. In works
such as \cite{GIMMSY-mm}, the former is used for obtaining the
{\sl Poincar\'e-Cartan} form in $J^1E$, which is needed for the
Lagrangian formalism. This is done by defining the suitable {\sl
Legendre map} connecting $J^1E$ and ${\cal M}\pi$. On the other
hand, the dimensions of $J^1E$ and ${\cal M}\pi$ and $J^1E^*$ are
not equal: in fact, $\dim\,{\cal M}\pi=\dim\, J^1E+1$ and
$\dim\,J^1E^*=\dim\, J^1E+(\dim\, M)^2$;
 but $\dim\,\Pi=\dim\, J^1\pi^*=\dim\, J^1E$
 and the choice of both $J^1\pi^*$
 and $\Pi$ as multimomentum phase spaces allows us to state
coherent covariant Hamiltonian formalisms for Field Theories.
Finally, the construction of $J^1\pi^*$ is closely related to
${\cal M}\pi$, but their relation to $\Pi$ and $J^1E^*$ is not
evident at all.

Hence, the aim of this work is to carry out a comparative study of
these multimomentum bundles, and introduce the canonical
geometrical structures of ${\cal M}\pi$ and $J^1E^*$. In every
case, the corresponding {\sl Legendre map} is also defined. An
interesting conclusion of this study is that the multimomentum
bundles $J^1\pi^*$ and $\Pi$ are canonically diffeomorphic.
Finally, using the different Legendre maps, we can classify
Lagrangian systems in Field Theory into {\sl (hyper)regular} and
{\sl almost-regular}, from different but equivalent ways,
attending to the characteristics of these maps.

All manifolds are real, paracompact, connected and $C^\infty$.
 All maps are $C^\infty$.
Sum over crossed repeated indices is understood.

\section{Multimomentum bundles}

 Let $\pi\colon E\to M$ be a fiber bundle ($\dim\, E=N+m$, $\dim\,
M=m$),
 $\pi^1\colon J^1E\to E$ the first order jet bundle of local sections
 of $\pi$, and $\bar\pi^1=\pi \circ \pi^1$.
 $J^1E$ is an affine bundle modeled on
 $\pi^*\Tan^*M\otimes{\rm V}(\pi)$.

A local chart of natural coordinates in $E$ adapted to the bundle
$E\to M$ will be denoted by $(x^\mu,y^A)$. The induced local chart
in $J^1E$ is denoted by $(x^\mu,y^A,v^A_\mu)$.

\begin{definition}
 The bundle (over $E$)
 $$
 J^1E^*:=\pi^*\Tan M\otimes_E\Tan^*E\otimes_E\pi^*\Lambda^m\Tan^*M
 $$
 is called the {\rm generalized multimomentum bundle} associated with
the bundle $\pi\colon E\to M$.
 We denote  the natural projections by $\hat\rho^1\colon J^1E^*\to E$
and
 $\bar{\hat\rho^1}:=\pi\circ\hat\rho^1\colon J^1E^*\to M$.

\end{definition}

 The local system $(x^\mu,y^A)$ induces a local
 system of natural coordinates
 $(x^\mu,y^A,{\rm p}_\mu^\nu,{\rm p}_A^\mu)$ in $J^1E^*$ as follows:
 if ${\bf y}\in J^1E^*$, with
 ${\bf y}\stackrel{\hat\rho^1}{\to}y\stackrel{\pi}{\to}x$,
 we have that
\beq
 {\bf y}=\derpar{}{x^\mu}\Big\vert_y\otimes
 (f^\mu_\nu\d x^\nu+g_A^\mu\d y^A)_y\otimes\d^mx\vert_y
\label{element}
\eeq
 (where $\d^mx\equiv\d x^1\wedge\ldots\wedge\d x^m$), and therefore
 $$
\begin{array}{ccc}
 x^\mu({\bf y})=x^\mu ((\pi\circ\hat\rho^1)({\bf y})) & ; &
 y^A({\bf y})=y^A(\hat\rho^1({\bf y}))
  \\
 {\rm p}^\mu_\nu({\bf y})=
 {\bf y}\left(\d x^\mu\otimes\derpar{}{x^\nu}\otimes\partial^mx
 \Big\vert_y\right) =f^\mu_\nu & ; &
 {\rm p}^\mu_A({\bf y})=
 {\bf y}\left(\d x^\mu\otimes\derpar{}{y^A}\otimes\partial^mx
 \Big\vert_y\right) =g^\mu_A
\end{array}
$$
 (where \dst\partial^mx\equiv\derpar{}{x^1}
 \wedge\ldots\wedge\derpar{}{x^m}\) ).

\begin{definition}
 The bundle (over $E$)
$$
\Pi:=\pi^*\Tan M\otimes_E{\rm V}^*(\pi )\otimes_E
\pi^*\Lambda^m\Tan^*M =
\bigcup_{y\in E} \Tan_{\pi (y)}M\otimes{\rm V}_y^*(\pi )\otimes
\Lambda^m\Tan_{\pi (y)}^*M
$$
 is called the {\rm reduced multimomentum bundle} associated with the
bundle $\pi\colon E\to M$.
 We denote the natural projections by
 $\rho^1\colon \Pi\to E$  and
 $\bar\rho^1:=\pi\circ\rho^1\colon \Pi\to M$.
\end{definition}

 From the local system $(x^\mu ,y^A)$
 we can construct a natural system of coordinates
 $(x^\mu ,y^A,{\rm p}_A^\mu )$ in $\Pi$ as follows: considering
 $\tilde y\in \Pi$, we write
$$
 x^\mu (\tilde y) = x^\mu (\bar\rho^1 (\tilde y)) \ ; \
 y^A (\tilde y) = y^A (\rho^1 (\tilde y)) \ ; \
 {\rm p}_A^\mu (\tilde y) = \tilde y\left(\d x^\mu ,\derpar{}{y^A},
 \derpar{}{x^1}\wedge\ldots\wedge\derpar{}{x^m}\right)
$$
 and denoting the dual basis of \dst\derpar{}{y^A}\)
 in ${\rm V}^*(\pi )$  by $\{\zeta^A\}$,
 an element $\tilde y\in \Pi$ is expressed as
$$
\tilde y =
 {\rm p}^\mu_A(\tilde y)\derpar{}{x^\mu}\otimes\zeta^A\otimes\d^mx
\Big\vert_{(x^\mu (\tilde y),y^A(\tilde y))}
$$

The relation between these multimomentum bundles is given by the (onto)
map
$$
\begin{array}{ccccc}
\delta &\colon& J^1E^* &\longrightarrow& \Pi
\\
& & (x^\mu,y^A,{\rm p}_A^\mu,{\rm p}_\mu^\nu)
 & \mapsto & (x^\mu,y^A,{\rm p}_A^\mu)
\end{array}
$$
which is induced by the natural restriction $\Tan^*E\to{\rm V}^*(\pi)$.

\begin{definition}
Consider the multicotangent bundle $\Lambda^m\Tan^*E$. Then,
for every $y\in E$ we define
$$
\Lambda_1^m\Tan_y^*E:=
\{\gamma\in\Lambda^m\Tan_y^*E \ ;\
\inn (u_1)\inn (u_2)\gamma =0\ ,\ u_1,u_2\in{\rm V}_y(\pi)\}
$$
The bundle (over $E$)
$$
 {\cal M}\pi\equiv\Lambda_1^m\Tan^*E :=
 \bigcup_{y\in E}\Lambda_1^m\Tan_y^*E=
 \bigcup_{y\in E} \{ (y,\alpha ) \ ;\ \alpha\in\Lambda_1^m\Tan_y^*E\}
$$
 will be called the {\rm extended multimomentum bundle}
 associated with the bundle $\pi\colon E\to M$.
 We denote  the natural projections by
  $\hat\kappa^1\colon{\cal M}\pi\to E$ and
 $\bar{\hat\kappa}^1\colon{\cal M}\pi\to M$.
\end{definition}

 The local chart $(x^\mu,y^A)$ in $E$ induces a natural system of
 coordinates $(x^\mu,y^A,p,p_A^\mu)$ in ${\cal M}\pi$. Hence, if
 ${\hat y}\in{\cal M}\pi$
 (with ${\hat y}\stackrel{\hat\kappa^1}{\to}y\stackrel{\pi}{\to}x$),
 it is a $m$-covector whose expressions in a natural chart is
 $$
 \hat y=\lambda\,\d^mx+\lambda^\mu_A\d y^A\wedge\d^{m-1}x_\mu
 $$
  and we have that
 $$
  x^\mu({\hat y})=x^\mu (y) \ ,\ y^A({\hat y})=y^A(y) \ ,\
  p({\hat y})=\hat y(\partial^mx)=\lambda \ ,\
  p^\mu_A({\hat y})=\hat y
 \left(\derpar{}{y^A}\wedge\partial^{m-1}x^\mu\right)
 =\lambda^\mu_A
 $$
  (where $\partial^{m-1}x^\mu\equiv\inn(\d x^\mu)\partial^mx$).

 Another usual characterization of the bundle
 ${\cal M}\pi$ is the following:

\begin{prop}
 ${\cal M}\pi\equiv\Lambda_1^m\Tan^*E$
 is canonically isomorphic to
 ${\rm Aff}(J^1E,\pi^*\Lambda^m\Tan^*M)$.
\label{propoaux}
\end{prop}
\proof  It is a consequence of Lemma \ref{isocan1} of the
appendix, taking $G=\Tan_yE$, $H=\Tan_xM$, $F={\rm V}_y(\pi)$, and
$\Sigma=J^1_yE$, and then observing that the sequence
(\ref{sequence}) is $$ 0\longrightarrow {\rm V}_y(\pi)
\stackrel{j_y}{\longrightarrow}
\Tan_yE\stackrel{\Tan_y\pi}{\longrightarrow}\Tan_xM\longrightarrow
0 $$ (See also \cite{CCI-91}).\qed

{\bf Comment}: \bit \item
Given a section
 ${\mit\Gamma}\colon E\to J^1E$ of $\pi^1$,
 in the same way as it is commented in the Lemma \ref{isocan}
 of the appendix, we have an splitting
 $$
{\rm Aff}\, (J^1E,\pi^*\Lambda^m\Tan^*M)\simeq
\pi^*\Lambda^m\Tan^*M\oplus (\pi^*\Lambda^m\Tan^*M\otimes {\rm
V}^*(\pi))
 $$
 and them $\dim\, ({\cal M}\pi)_y=\dim\, \Pi_y+1$,
 for every $y\in E$.
  \eit

 We can introduce the {\sl canonical contraction} in $J^1E^*$,
which is defined as the map $$
\begin{array}{ccccc}
\iota & \colon & J^1E^*\equiv\pi^*\Tan
M\otimes\Tan^*E\otimes\pi^*\Lambda^m\Tan^*M & \longrightarrow &
\Lambda^m\Tan^*E \\ & & {\bf y}=u_k\otimes\alpha^k \otimes\chi &
\mapsto & \alpha^k\wedge\pi^*\inn (u_k)\chi
\end{array}
$$ In a natural chart $(x^\mu,y^A,{\rm p}_\mu^\nu,{\rm p}_A^\mu)$
in $J^1E^*$, bearing in mind (\ref{element}), we obtain $$ \iota
({\bf y}) = (f^\mu_\nu\d x^\nu+g_A^\mu\d y^A)_y\wedge
\inn\left(\derpar{}{x^\mu}\right)\d^mx\Big\vert_y=
(f^\mu_\mu\d^mx+g_A^\mu\d y^A\wedge\d^{m-1}x_\mu)_y $$ (where
\dst\d^{m-1}x_\mu\equiv\inn\left(\derpar{}{x^\mu}\right)\d^mx\) .
Remember that $f^\mu_\mu$ denotes \dst\sum_{\mu=1}^mf^\mu_\mu\) ).

Therefore, the relation between the multimomentum bundles
${\cal M}\pi$ and $J^1E^*$ is:

\begin{prop}
${\cal M}\pi =\iota (J^1E^*)$. (We will denote $\iota_0\colon
J^1E^*\to {\cal M}\pi$ the restriction of $\iota$ onto its image
${\cal M}\pi$).
\end{prop}
\proof For every $y\in E$ we must prove that $$ \iota (J^1E^*)_y
=\{\gamma\in\Lambda^m\Tan_y^*E \ ;\ \inn (u_1)\inn (u_2)\gamma =0\
,\ u_1,u_2\in{\rm V}_y(\pi)\} \equiv \Lambda_1^m\Tan_y^*E $$ In
fact; if ${\bf y}\in J^1E^*$, $\gamma =\iota ({\bf y})$ and
$u_1,u_2\in{\rm V}_y\pi$, in a local chart we have $$
\inn(u_1)\inn(u_2)(\iota({\bf y}))= \inn(u_1)\inn(u_2)({\rm
p}^\nu_\nu({\bf y})\d^mx+p^\mu_A({\bf y})\d
y^A\wedge\d^{m-1}x_\mu)_y=0 $$ and conversely, if
$\gamma\in\Lambda^m\Tan_y^*E$ satisfies the above condition, then
$\gamma =(\lambda\d^mx+\lambda^\mu_A\d y^A\wedge\d^{m-1}x_\mu)_y$,
therefore $\gamma=\iota({\bf y})$, with $$ {\bf
y}=\derpar{}{x^\mu}\Big\vert_y\otimes
\left(\left(\frac{\lambda}{m}(\d x^1+\ldots +\d
x^m)+\lambda^\mu_A\d y^A\right)\Big\vert_y
\otimes\d^mx\vert_y\right) $$ (Observe that $\Lambda_1^m\Tan^*E$
is canonically isomorphic to
$\Tan^*E\wedge\pi^*\Lambda^{m-1}\Tan^*M$). \qed

Note that, in natural coordinates, we have
$$
\iota\colon {\bf y}\equiv (x^\mu,y^A,{\rm p}_A^\mu,{\rm p}_\mu^\nu)
\mapsto \hat y\equiv (x^\mu,y^A,{\rm p}_A^\mu,{\rm p}=p_\mu^\mu)
$$

The sections of the bundle
$\pi^*\Lambda^m\Tan^*M\to E$ are the $\pi$-semibasic $m$-forms
on $E$. Therefore we introduce the notation
$\Lambda_0^m\Tan^*E\equiv\pi^*\Lambda^m\Tan^*M$, and then:

\begin{definition}
The bundle (over $E$) $$ J^1\pi^*:=
\Lambda_1^m\Tan^*E/\Lambda_0^m\Tan^*E\equiv{\cal
M}\pi/\Lambda_0^m\Tan^*E $$ will be called the {\rm restricted
multimomentum bundle} associated with the bundle $\pi\colon E\to
M$. We denote  the natural projections by $\kappa^1\colon
J^1\pi^*\to E$ and $\bar\kappa^1:=\pi\circ\kappa^1\colon
J^1\pi^*\to M$.
\end{definition}

The natural coordinates
in $J^1\pi^*$ will be denoted as $(x^\mu,y^A,p_A^\mu)$.

The relation between the bundles ${\cal M}\pi$ and $J^1\pi^*$ is given
by
the natural projection
$$
\begin{array}{ccccc}
\mu &\colon& {\cal M}\pi=\Lambda_1^m\Tan^*E &\longrightarrow&
\Lambda_1^m\Tan^*E/\Lambda_0^m\Tan^*E=J^1\pi^*
\\
& & (x^\mu,y^A,p_A^\mu,p) & \mapsto & (x^\mu,y^A,p_A^\mu)
\end{array}
$$
Finally, the relation between the multimomentum bundles $\Pi$ and
$J^1\pi^*$ is:

\begin{teor}
The multimomentum bundles $J^1\pi^*$ and $\Pi$ are canonically
diffeomorphic. We will denote this diffeomorphism by
${\mit\Psi}\colon J^1\pi^*\to\Pi$. \label{teoisom}
\end{teor}
\proof By Proposition \ref{propoaux}, we have that
$\Lambda_1^m\Tan^*E\equiv {\cal M}\pi$ is canonically isomorphic
to ${\rm Aff}(J^1E,\Lambda^m\Tan^*M)$. On the other hand, taking
$G=\Tan_yE$, $H=\Tan_xM$, $F={\rm V}_y(\pi)$, and $\Sigma=J^1_yE$
in Lemma \ref{isocan2} of the appendix, we obtain $$ {\rm
Aff}(J^1_yE,\Lambda^m\Tan_x^*M)/\Lambda^m\Tan_x^*M \ \simeq\
\Tan_xM\otimes{\rm V}_y^*(\pi)\otimes\Lambda^m\Tan_x^*M $$
Therefore, extending these constructions to the bundles we have $$
J^1\pi^*\ \simeq\ {\rm
Aff}(J^1E,\pi^*\Lambda^m\Tan^*M)/\pi^*\Lambda^m\Tan^*M \ \simeq\
\pi^*\Tan M\otimes{\rm V}^*(\pi)\otimes\pi^*\Lambda^m\Tan^*M :=
\Pi $$ \qed

{\bf Remark}:
\begin{itemize}
  \item
As is known, a connection in the bundle $\pi\colon E\to M$, that
is, a section ${\mit\Gamma}\colon E\to J^1E$ of $\pi^1$, induces a
linear map $\bar{\mit\Gamma}\colon {\rm V}^*(\pi)\to \Tan^*E$ and,
as a consequence, another one
 $$ \begin{array}{ccccc}
 \tilde{\mit\Gamma}&\colon&
 \pi^*\Tan M\otimes{\rm V}^*(\pi)\otimes\Lambda^m\Tan^*M
  &\to & \Lambda^m_1\Tan^*E
  \\
  & & u\otimes\alpha\otimes\xi &\mapsto &
  \bar{\mit\Gamma}(\alpha)\wedge\inn(u)\xi
  \end{array} $$
  In this way, we can get the inverse map ${\mit\Psi}^{-1}$
  by means of a connection: it is the composition of
  $\tilde{\mit\Gamma}$ with
  $\mu\colon\Lambda_1^m\Tan^*E\to\Lambda_1^m\Tan^*E/\Lambda_0^m\Tan^*E$.
  Nevertheless, ${\mit\Psi}^{-1}$ is connection independent
  because, given two connections ${\mit\Gamma}_1,{\mit\Gamma}_2$,
  the image of  ${\mit\Gamma}_1-{\mit\Gamma}_2$ is in
  $\Lambda_0^m\Tan^*E\subset\Lambda_1^m\Tan^*E$.
\end{itemize}

 Next, we give the coordinate expression
of this diffeomorphism. First, let us recall that the natural
coordinates in $E$ produce an affine reference frame in $J^1E$ as
follows: if $\pi (y)=x$, take $\bar y_0,\bar y^\mu_A\in J^1_yE$ as
\beann \bar y_0 &=& \{\phi\colon M\to E \ ;\ \phi (x)=y \ , \
\Tan_x\phi\left(\derpar{}{x^\mu}\right)_x=
\left(\derpar{}{x^\mu}\right)_y\}
\\
\bar y^\mu_A &=& \{\phi\colon M\to E \ ;\ \phi (x)=y \ , \
\Tan_x\phi\left( \derpar{}{x^\mu}\right)_x=
\derpar{}{x^\mu}\Big\vert_y+\derpar{}{y^A}\Big\vert_y \} \eeann
(Observe that $\bar y_0$ is the 1-jet of critical sections with
target $y$); then we obtain $\bar y-\bar y_0 =v^A_\mu(\bar y)(\bar
y^\mu_A-\bar y_0)$, for every $\bar y\in J^1_yE$.

An affine map $\varphi\colon J^1E\to\Lambda^m_0\Tan^*E$ is given
by $\varphi=(\varphi(\bar y_0),\hat\varphi)$ (where $\hat\varphi$
denotes the linear part of $\varphi$), then
$$
 \varphi (\bar
y)=\varphi (\bar y_0)+\hat\varphi(v^A_\mu(\bar y)(\bar
y^\mu_A-\bar y_0))= \varphi (\bar y_0)+v^A_\mu(\bar
y)\hat\varphi(\bar y^\mu_A-\bar y_0)
$$
 Denoting by $q$ the fiber
coordinate in $\pi^*\Lambda^m\Tan^*M\equiv\Lambda^m_0\Tan^*E$. We
have $$ q(\varphi (\bar y))= (\varphi (\bar y_0))(\partial^mx)+
v^A_\mu(\bar y)\hat\varphi(\bar y^\mu_A-\bar y_0)(\partial^mx)=
\lambda+v^A_\mu(\bar y)\lambda^\mu_A $$ Then we have an affine
coordinate system in ${\rm Aff}(J^1E,\Lambda^m_0\Tan^*E)$, denoted
$(q,q^\mu_A)$, with $$ q(\varphi)=q(\varphi(\bar y_0)) \quad ,
\quad q^\mu_A(\varphi)=q(\hat\varphi(\bar y^\mu_A-\bar y_0)) $$
and hence $\varphi=(l,l^\mu_A)$, with $l=q(\varphi)$,
$l^\mu_A=q^\mu_A(\varphi)$ or, what means the same thing,
$\varphi(v^A_\mu)=(l+l^\mu_Av^A_\mu)\d^mx$.

Now, let $\bar y\in J^1E$, with $\bar
y\stackrel{\pi^1}{\to}y\stackrel{\pi}{\to}x$.
Consider the map
$$
\begin{array}{ccccccc}
\Upsilon&\colon&\Lambda_1^m\Tan^*E&\longrightarrow&{\rm
Aff}(J^1E,\Lambda^m_0\Tan^*E)& &
\\
& & \eta & \mapsto & \Upsilon (\eta) & \colon & \bar y \mapsto
(\phi^*\eta)_y
\end{array}
$$
 (introduced in Lemma \ref{isocan1} of the appendix),
 where $\phi\colon M\to E$ is a representative of $\bar y$,
with $\phi(x)=y$. If $\beta= (\lambda\d^mx+\lambda^\mu_A\d
y^A\wedge\d^{m-1}x_\mu)_y\in\Lambda^m_1\Tan^*E$,
then
$$
\Tan_x\phi=\left(\matrix{{\rm Id} \cr v^A_\mu(y)\cr}\right) \quad
,\quad \Tan_x\phi\left( \derpar{}{x^\mu}\right)_x=
\left(\derpar{}{x^\mu}\right)_y+vÂ_\mu(\bar
y)\left(\derpar{}{y^A}\right)_y
$$
 Therefore we obtain
 $$
(\Upsilon(\beta))(\bar y)=(\phi^*\beta)_y=
((\lambda+\lambda^\mu_Av^A_\mu(\bar y))\d^mx)_y
$$
 that is,
$\Upsilon(\beta)=(\lambda,\lambda^\mu_A)$ or, what is equivalent,
$$
 q(\Upsilon(\beta))=p(\beta)=\lambda \ , \
q^\mu_A(\Upsilon(\beta))=p^\mu_A(\beta)=\lambda^\mu_A \quad {\rm
or} \quad \Upsilon^*q=p \ , \ \Upsilon^*q^\mu_A=p^\mu_A $$
 Thus $\Upsilon$ is a diffeomorphism in the fiber,
 and therefore a diffeomorphism, since it is the identity on the base.

Finally, we have the (commutative) diagram $$
\begin{array}{ccc}
\Lambda^m_1\Tan^*E & $\rightarrowfill$ & {\rm
Aff}(J^1E,\Lambda^m_0\Tan^*E)
\\
& \Upsilon &
\\
\Big\downarrow &
&\Big\downarrow
\\
& {\mit\Psi} &
\\
\Lambda^m_1\Tan^*E/\Lambda^m_0\Tan^*E  & $\rightarrowfill$ &
{\rm Aff}(J^1E,\Lambda^m_0\Tan^*E)/\Lambda^m_0\Tan^*E
\end{array}
$$
 and, for proving that $\Upsilon$ goes to the quotient,
 it suffices to see that
 $\Upsilon(\Lambda^m_0\Tan^*E)\subset\Lambda^m_0\Tan^*E$. To show
 this we must identify $\Lambda^m_0\Tan^*E$ as a subset of both
 $\Lambda^m_1\Tan^*E$ and ${\rm Aff}(J^1E,\Lambda^m_0\Tan^*E)$.
 In $\Lambda^m_1\Tan^*E$ we have that the elements of
 $\Lambda^m_0\Tan^*E$ are characterized as follows:
 $\beta\in\Lambda^m_0\Tan^*E$ iff, for a natural
 coordinate system $(x^\mu,y^A,p,p^\mu_A)$, we have
 $p^\mu_A(\beta)=0$. On the other hand, as a subset of
  ${\rm  Aff}(J^1E,\Lambda^m_0\Tan^*E)$,
  $$
  \Lambda^m_0\Tan^*E \equiv
 \{\varphi\in {\rm Aff}(J^1E,\Lambda^m_0\Tan^*E)\ ;
 \ \varphi={\rm constant}\} =
  \{\varphi\in {\rm Aff}(J^1E,\Lambda^m_0\Tan^*E)
   \ ;\  \hat\varphi=0 \}
$$
 ($\hat\varphi$ denotes the linear part of $\varphi$),
 or equivalently, for an affine natural coordinate system
 $(q,q^\mu_A)$,
 $$
  \Lambda^m_0\Tan^*E\equiv
  \{\varphi\in {\rm Aff}(J^1E,\Lambda^m_0\Tan^*E) \ ;\
  q^\mu_A(\varphi)=0\}
 $$
 And, from the local expression of $\Upsilon$, we obtain that
 $\Upsilon(\Lambda^m_0\Tan^*E)=\Lambda^m_0\Tan^*E$. As a
consequence of this, if $p^\mu_A$ are the fiber coordinates in
$J^1\pi^*=\Lambda^m_1\Tan^*E/\Lambda^m_0\Tan^*E$, and ${\rm
p}^\mu_A$ are those in $\Pi={\rm
Aff}(J^1E,\Lambda^m_0\Tan^*E)/\Lambda^m_0\Tan^*E$, we have that
 $$
{\mit\Psi}^*{\rm p}^\mu_A=p^\mu_A \quad ,\quad \forall \mu,A
 $$
 and, consequently, in these natural coordinate systems,
 the diffeomorphism ${\mit\Psi}$ is the identity.

\section{Canonical forms in the multimomentum bundles}

The multimomentum bundles $J^1E^*$ and ${\cal M}\pi$ are endowed with
canonical differential forms:

\begin{definition}
The {\rm canonical $m$-form} of $J^1E^*$ is the form
$\hat\Theta\in\df^m(J^1E^*)$ defined as follows: if ${\bf y}\in
J^1E^*$ and $\moment{X}{1}{m}\in\vf (J^1E^*)$, then $$ \hat\Theta
({\bf y};\moment{X}{1}{m}) := \iota({\bf y}) (\Tan_{{\bf
y}}\hat\rho^1(X_1),\ldots ,\Tan_{{\bf y}}\hat\rho^1(X_m)) $$

The {\rm canonical $(m+1)$-form} of $J^1E^*$ is
$\hat\Omega :=-\d\hat\Theta\in\df^{m+1}(J^1E^*)$.
\end{definition}

In a natural chart in $J^1E^*$ we have
\beann
\hat\Theta &=& {\rm p}^\nu_\nu\d^mx+{\rm p}^\mu_A\d
y^A\wedge\d^{m-1}x_\mu
\\
\hat\Omega &=&
-\d {\rm p}^\nu_\nu\wedge\d^mx-\d {\rm p}^\mu_A\wedge\d
y^A\wedge\d^{m-1}x_\mu
\eeann

{\bf Remark}:
 \bit
\item
As is known \cite{CIL-98}, the {\sl multicotangent bundle}
$\Lambda^m\Tan^*E$ is endowed with canonical forms: ${\bf
\Theta}\in\df^m(\Lambda^m\Tan^*E)$ and the multisymplectic form
${\bf  \Omega}:=-\d{\bf \Theta}\in\df^{m+1}(\Lambda^m\Tan^*E)$.
Then, it can be easily proved that $\hat\Theta=\iota^*{\bf
\Theta}$. \eit

Observe that ${\cal M}\pi\equiv\Lambda^m_1\Tan^*E$ is a subbundle
of the {\sl multicotangent bundle} $\Lambda^m\Tan^*E$. Let
$\varsigma\colon\Lambda^m_1\Tan^*E\hookrightarrow\Lambda^m\Tan^*E$
be the natural imbedding (hence $\varsigma\circ\iota_0=\iota$).
Then:

\begin{definition}
The {\rm canonical $m$-form} of ${\cal M}\pi$ is
$\Theta :=\varsigma^* {\bf \Theta}\in\df^m({\cal M}\pi)$.
It is also called the {\rm multimomentum Liouville $m$-form}.

The {\rm canonical $(m+1)$-form} of ${\cal M}\pi$ is
$\Omega =-\d\Theta\in\df^{m+1}({\cal M}\pi)$,
and it is called the {\rm multimomentum Liouville $(m+1)$-form}.
\end{definition}

The expressions of these forms in a natural chart in ${\cal M}\pi$ are
\beann
\Theta&=&p\d^mx+p^\mu_A\d y^A\wedge\d^{m-1}x_\mu
\\
\Omega&=&-\d p\wedge\d^mx-\d p^\mu_A\wedge\d
y^A\wedge\d^{m-1}x_\mu \eeann Then, a simple calculation allows us
to prove that $\Omega$ is $1$-nondegenerate, and hence $({\cal
M}\pi,\Omega )$ is a {\sl multisymplectic manifold}.

{\bf Remark}:
 \bit \item
 Considering the natural projection
$\hat\kappa^1\colon\Lambda^m_1\Tan^*E\to E$, then $$ \Theta
((y,\alpha );\moment{X}{1}{m}):= \alpha
(y;\Tan_{(y,\alpha)}\hat\kappa^1(X_1),\ldots ,\Tan_{(y,\alpha
)}\hat\kappa^1 (X_m)) $$ for every $(y,\alpha
)\in\Lambda^m_1\Tan^*E$ (where $y\in E$ and
$\alpha\in\Lambda^m_1\Tan_y^*E$), and $X_i\in\vf
(\Lambda^m_1\Tan^*E)$. \eit

 The canonical $m$-forms $\hat\Theta$
and $\Theta$ can be characterized alternatively in the following
way:

\begin{prop}
\ben
\item
$\hat\Theta$ is the unique $\hat\rho^1$-semibasic $m$-form in $J^1E^*$
such that,
for every section $\hat\psi\colon E\to J^1E^*$ of $\hat\rho$,
the relation $\hat\psi^*\hat\Theta =\iota\circ\hat\psi$ holds.
\item
$\Theta$ is the only $\hat\kappa^1$-semibasic $m$-form
in ${\cal M}\pi$ such that,
for every section $\phi\colon E\to\Lambda^m_1\Tan^*E$ of $\hat\kappa^1$,

the relation $\phi^*\Theta =\phi$ holds.
\een
\end{prop}
\proof
\ben
\item
  From the definition of $\hat\Theta$ it is obvious that it is
$\hat\rho^1$-semibasic. Then, for the second relation,
let $y\in E$ and $\moment{u}{1}{m}\in\Tan_yE$; therefore
\beann
(\hat\psi^*\hat\Theta)(y;\moment{u}{1}{m}) &=&
\hat\Theta(\hat\psi(y);\Tan_y\hat\psi(u_1),\ldots ,\Tan_y\hat\psi(u_m))
\\ &=&
\iota(\hat\psi(y))
(y;(\Tan_{\hat\psi(y)}\hat\rho^1\circ\Tan_y\hat\psi)(u_1),
\ldots ,(\Tan_{\hat\psi(y)}\hat\rho^1\circ\Tan_y\hat\psi)(u_m))
\\ &=&
\iota(\hat\psi(y))(y;\moment{u}{1}{m}) =
(\iota\circ\hat\psi)(y;\moment{u}{1}{m})
\eeann
Conversely, suppose that $\hat\Theta\in\df^m(J^1E^*)$ verifies both
conditions in the statement. We will prove that
$\hat\Theta$ is uniquely determined. Let ${\bf y}\in J^1E^*$,
with ${\bf y}\stackrel{\hat\rho^1}{\to}y\stackrel{\pi}{\to}x$,
and $\moment{v}{1}{m}\in\Tan_{{\bf y}}J^1E^*$.
We can suppose that $\moment{v}{1}{m}$ are linearly independent and that

$\langle\moment{v}{1}{m}\rangle\cap{\rm V}_{\bar
y}(\pi\circ\hat\rho^1)=\{ 0\}$ (where
$\langle\moment{v}{1}{m}\rangle$ denotes the subspace generated by
these vectors), for in other case $\hat\Theta ({\bf
y};\moment{v}{1}{m})=0$. Then, by the Lemma \ref{lema0} (see the
appendix), there exist $\moment{u}{1}{m}\in\Tan_xM$ and a local
section $\hat\psi\colon M\to J^1E^*$ of $\pi\circ\hat\rho^1$, such
that $\hat\psi(x)={\bf y}$ and $\Tan_x\hat\psi (u_\mu)=v_\mu$
($\mu=1,\ldots ,m$), and we have that \beann \hat\Theta ({\bf
y};\moment{v}{1}{m}) &=& \hat\Theta
(\hat\psi(y);\Tan_y\hat\psi(u_1), \ldots ,\Tan_y\hat\psi(u_m))
\\ &=&
(\hat\psi^*\hat\Theta)(y;\moment{u}{1}{m}) =
(\iota\circ\hat\psi)(y;\moment{u}{1}{m})
\eeann
Hence $\hat\Theta ({\bf y};\moment{v}{1}{m})$ is uniquely determined.
\item
The proof follows the same pattern as the one above.
 \een \qed

\section{Legendre maps}
 \protect\label{lm}

 From the Lagrangian point of view,
a {\sl classical Field Theory} is described by its {\sl
configuration bundle} $\pi\colon E\to M$ (where $M$ is an oriented
manifold with volume form $\omega\in\df^m(M)$); and a {\sl
Lagrangian density} which is a $\bar\pi^1$-semibasic $m$-form on
$J^1E$. A Lagrangian density is usually written as $\Lag =\lag
(\bar\pi^{1^*}\omega)$, where $\lag\in\Cinfty (J^1E)$ is the {\sl
Lagrangian function} associated with $\Lag$ and $\omega$. Then a
{\sl Lagrangian system} is a couple $\ls$. The {\sl
Poincar\'e-Cartan $m$ and $(m+1)$-forms} associated with the
Lagrangian density $\Lag$ are defined using the {\sl vertical
endomorphism} ${\cal V}$ of the bundle $J^1E$:
 $$
\Theta_{\Lag}:=\inn({\cal
V})\Lag+\Lag\equiv\theta_{\Lag}+\Lag\in\df^{m}(J^1E) \quad ;\quad
\Omega_{\Lag}:= -\d\Theta_{\Lag}\in\df^{m+1}(J^1E) $$
 In a natural chart in $J^1E$, in which
 $\bar\pi^{1^*}\omega=\d^mx$, we have
\beann \Theta_{\Lag}&=&\derpar{\lag}{v^A_\mu}\d
y^A\wedge\d^{m-1}x_\mu - \left(\derpar{\lag}{v^A_\mu}v^A_\mu
-\lag\right)\d^mx
\\
\Omega_{\Lag}&=& -\frac{\partial^2\lag}{\partial v^B_\nu\partial
v^A_\mu} \d v^B_\nu\wedge\d y^A\wedge\d^{m-1}x_\mu
-\frac{\partial^2\lag}{\partial y^B\partial v^A_\mu}\d y^B\wedge
\d y^A\wedge\d^{m-1}x_\mu + \nonumber \\ & &
\frac{\partial^2\lag}{\partial v^B_\nu\partial v^A_\mu}v^A_\mu \d
v^B_\nu\wedge\d^mx  + \left(\frac{\partial^2\lag}{\partial
y^B\partial v^A_\mu}v^A_\mu -\derpar{\lag}{y^B}+
\frac{\partial^2\lag}{\partial x^\mu\partial v^B_\mu} \right)\d
y^B\wedge\d^mx \eeann (For a more detailed description of all
these concepts see, for instance, \cite{BSF-88}, \cite{EMR-96},
\cite{Gc-73}, \cite{GS-73}, \cite{Sa-89}).

For constructing the Hamiltonian formalism associated with a
Lagrangian system in Field Theory, the {\sl Legendre maps} are
introduced. Then, depending on the choice of the multimomentum
bundle, we can define different types of these maps, as follows:

\begin{definition}
Let $\ls$ be a Lagrangian system,
and $\bar y\in J^1E$,
with $\bar y\stackrel{\pi^1}{\mapsto}y\stackrel{\pi}{\mapsto}x$.
\ben
\item
Let ${\cal D}\subset\Tan J^1E$ be the subbundle
 of total derivatives in $J^1E$ (which in a system of natural
 coordinates in $J^1E$, is generated by
\dst\left\{\derpar{}{x^\mu}+v^A_\mu\derpar{}{y^A}\right\}\) ). We
have that $\pi^{1*}\Tan E=\pi^{1*}{\rm V}(\pi)\oplus{\cal D}$ with
$\Tan_yE\big\vert_{\bar y}={\rm V}_y\pi\big\vert_{\bar
y}\oplus{\cal D}_{\bar y}$ (see \cite{Sa-89} for details). Hence
there is a natural projection $\sigma\colon\pi^{1*}\Tan
E\to\pi^{1*}{\rm V}(\pi)$, and we can draw the diagram $$
\begin{array}{cccc}
\begin{picture}(80,60)(0,0)
\put(0,49){\mbox{$\Tan_{\bar y}J^1_yE={\rm V}_{\bar y}\pi^1\simeq$}}
\end{picture}&
\begin{picture}(70,60)(0,0)
\put(0,49){\mbox{$(\Tan_x^*M\otimes{\rm V}_y\pi)_{\bar y}$}}
\put(50,25){\mbox{${\rm Id}\otimes\sigma_{\bar y}$}}
\put(35,15){\vector(0,1){25}}
\put(0,0){\mbox{$(\Tan_x^*M\otimes\Tan_yE)_{\bar y}$}}
\end{picture}
&
\begin{picture}(70,60)(0,0)
\put(0,50){\vector(1,0){70}}
\put(25,55){\mbox{$D_{\bar y}\Lag_y$}}
\put(0,3){\vector(2,1){70}}
\end{picture}
&
\begin{picture}(20,60)(0,0)
\put(0,49){\mbox{$(\Lambda^m\Tan_x^*M)_{\bar y}$}}
\end{picture}
\end{array}
$$
Then, the {\rm generalized Legendre map} is the $\Cinfty$-map
$$
\begin{array}{ccccc}
\widehat{{\rm F}\Lag} & \colon & J^1E &$\rightarrowfill$ & J^1E^*
\\
& &\bar y & \mapsto & D_{\bar y}\Lag_y\circ ({\rm
Id}\otimes\sigma)_{\bar y}
\end{array}
$$
\item
The {\rm reduced Legendre map}  is the $\Cinfty$-map
$$
\begin{array}{ccccc}
{\rm F}\Lag & \colon & J^1E &$\rightarrowfill$ & \Pi
\\
& &\bar y & \mapsto & \tilde\Tan_{\bar y}\Lag_y
\end{array}
$$
where the map $\tilde\Tan_{\bar y}\Lag_y$ is defined from the
following diagram (where the vertical arrows are
canonical isomorphisms given by the directional derivatives)
$$
\begin{array}{ccc}
\Tan_{\bar y}J^1_yE & $\rightarrowfill$ &
\Tan_{\Lag_y(\bar y )}\Lambda^m\Tan_x^*M
\\
& \Tan_{\bar y}\Lag_y &
\\
\simeq\ \Big\updownarrow & & \Big\updownarrow\ \simeq
\\
& \tilde\Tan_{\bar y}\Lag_y &
\\
\Tan_x^*M\otimes{\rm V}_y\pi & $\rightarrowfill$ &
\Lambda^m\Tan_x^*M
\end{array}
$$
(${\rm F}\Lag$ is just the vertical derivative of $\Lag$ \cite{XGr-97}).

\item
The {\rm (first) extended Legendre map}  is the $\Cinfty$-map
$\widehat{{\cal F}\Lag}\colon J^1E \to {\cal M}\pi$ given by $$
\widehat{{\cal F}\Lag}:=\iota_0\circ\widehat{{\rm F}\Lag} $$ The
{\rm (second) extended Legendre map} is the $\Cinfty$-map
$\widetilde{{\cal F}\Lag}\colon J^1E \to {\cal M}\pi$ given by $$
\widetilde{{\cal F}\Lag}=\widehat{{\cal F}\Lag}+\pi^*\Lag $$
\item
The {\rm restricted Legendre map}  is
the $\Cinfty$-map ${\cal F}\Lag\colon J^1E \to \Pi$ given by
$$
{\cal F}\Lag:=\mu\circ\widehat{{\cal F}\Lag}=\mu\circ\widetilde{{\cal
F}\Lag}
$$
\een
\end{definition}

If $\bar y\in J^1E$, with $\pi^1(\bar y)=y$, using the natural
coordinates in the different multimomentum bundles, we have:
 \beann
 {\cal F}\Lag (\bar y)&=&
\derpar{\lag}{v^A_\mu}(\bar y) \d y^A
\wedge\d^{m-1}x_\mu\Big\vert_y
\\
\widetilde{{\cal F}\Lag} (\bar y)&=&
 \derpar{\lag}{v^A_\mu}(\bar y)\d y^A \wedge\d^{m-1}x_\mu\Big\vert_y
 +(\lag -v^A_\nu)(\bar y)
\derpar{\lag}{v^A_\mu}(\bar y)\d^mx\Big\vert_y
 \\
 \widehat{{\cal F}\Lag} (\bar y) &=&
 \derpar{\lag}{v^A_\mu}(\bar y) \d y^A
 \wedge\d^{m-1}x_\mu\Big\vert_y -v^A_\nu(\bar y)
 \derpar{\lag}{v^A_\mu}(\bar y)\d^mx\Big\vert_y
 \\
 \widehat{{\rm F}\Lag}(\bar y)&=&
\derpar{\lag}{v^A_\mu}(\bar y) \left(\derpar{}{x^\mu}\otimes (\d
y^A-v^A_\nu(\bar y)\d
 x^\nu)\otimes\d^mx\right)\Big\vert_y
\\
 {\rm F}\Lag (\bar y)&=&
 \derpar{\lag}{v^A_\mu}(\bar y)
\left(\derpar{}{x^\mu}\otimes\zeta^A\otimes\d^mx\right)
\Big\vert_y
 \eeann
that is, the local expressions of the Legendre maps are the
following:
 $$
\begin{array}{ccccccc}
 {\cal F}\Lag^*x^\mu = x^\mu &\ , \
 & {\cal F}\Lag^*y^A = y^A &\  , \ & {\cal F}\Lag^*p_A^\mu =
\derpar{\lag}{v^A_\mu} & &
\\
\widetilde{{\cal F}\Lag}^*x^\mu = x^\mu &\  ,\ & \widetilde{{\cal
 F}\Lag}^*y^A = y^A &\  , \ & \widetilde{{\cal F}\Lag}^*p_A^\mu =
\derpar{\lag}{v^A_\mu}  &\  , \ & \widetilde{{\cal F}\Lag}^*p =
\lag -v^A_\mu\derpar{\lag}{v^A_\mu}
\\
\widehat{{\cal F}\Lag}^*x^\mu = x^\mu &\  ,\ & \widehat{{\cal
 F}\Lag}^*y^A = y^A &\  , \ & \widehat{{\cal F}\Lag}^*p_A^\mu =
 \derpar{\lag}{v^A_\mu}  &\  , \ & \widehat{{\cal F}\Lag}^*p =
 -v^A_\mu\derpar{\lag}{v^A_\mu}
 \\
\widehat{{\rm F}\Lag}^*x^\mu = x^\mu &\  ,\ &
\widehat{{\rm F}\Lag}^*y^A = y^A &\  ,\ &
\widehat{{\rm F}\Lag}^*{\rm p}^\mu_A = \derpar{\lag}{v^A_\mu} &\  ,\ &
\widehat{{\rm F}\Lag}^*{\rm p}^\mu_\nu = -v^A_\nu\derpar{\lag}{v^A_\mu}
 \\
 {\rm F}\Lag^*x^\mu = x^\mu &\  , \ &
 {\rm F}\Lag^*y^A = y^A &\  , \ &
 {\rm F}\Lag^*{\rm p}_A^\mu = \derpar{\lag}{v^A_\mu} & &
 \end{array}
$$

{\bf Remarks}:
 \bit
\item
 Taking into account all the above results, it is immediate to
 prove that ${\rm F}\Lag=\delta\circ\widehat{{\rm F}\Lag}$
 (see diagrams (\ref{diag1}) and (\ref{diag2})), and
 ${\rm F}\Lag={\mit\Psi}\circ{\cal F}\Lag$.
\item
It is interesting to point out that, as $\Theta_{\Lag}$ and
$\theta_{\Lag}$
can be thought as $m$-forms on $E$ along the projection
$\pi^1\colon J^1E\to E$, the extended Legendre maps can be defined as
\beann
(\widehat{{\cal F}\Lag}(\bar y))(\moment{Z}{1}{m})&=&
(\theta_{\Lag})_{\bar y}(\moment{\bar Z}{1}{m})
\\
(\widetilde{{\cal F}\Lag}(\bar y))(\moment{Z}{1}{m})&=&
(\Theta_{\Lag})_{\bar y}(\moment{\bar Z}{1}{m})
\eeann
where $\moment{Z}{1}{m}\in\Tan_{\pi^1(\bar y)}E$,
and $\moment{\bar Z}{1}{m}\in\Tan_{\bar y}J^1E$ are such that
$\Tan_{\bar y}\pi^1\bar Z_\mu=Z_\mu$.

In addition, the (second) extended Legendre map can also be defined
as the ``first order  vertical Taylor approximation to $\lag$''
\cite{CCI-91}, \cite{GIMMSY-mm}.
\eit

 Finally, we have the following relations between the
 Legendre maps and the Poincar\'e-Cartan $(m+1)$-form in $J^1E$
 (which can be easily proved using natural systems
 of coordinates and the expressions of the Legendre maps):

\begin{prop}
Let $\ls$ be a Lagrangian system. Then:
 \beann
 \widetilde{{\cal F}\Lag}^*\Theta=\Theta_{\Lag}
 &\quad ;\quad&
\widetilde{{\cal F}\Lag}^*\Omega=\Omega_{\Lag}
 \\
  \widehat{{\cal F}\Lag}^*\Theta =
 \Theta_{\Lag}-\Lag =\theta_{\Lag}
 &\quad ;\quad&
\widehat{{\cal F}\Lag}^*\Omega=\Omega_{\Lag}-\d\Lag=-\d\theta_\Lag
\\
 \widehat{{\rm F}\Lag}^*\hat\Theta=\Theta_{\Lag}-\Lag =\theta_{\Lag}
 &\quad ;\quad&
\widehat{{\rm F}\Lag}^*\hat\Omega =
 \Omega_{\Lag}-\d\Lag=-\d\theta_\Lag
 \eeann
\end{prop}

 {\bf Remark}:
 \bit
 \item
 Observe that the Hamiltonian formalism is
 essentially the dual formalism of the Lagrangian model, by means
 of the Lagrangian density. Then, as $J^1E$ is an affine bundle, its
 affine dual can be identified with
 ${\rm Aff}\, (J^1E,\pi^*\Lambda^m\Tan^*M)\simeq{\cal M}\pi$,
 whose dimension is greater than $\dim\,J^1E$, and hence
 ${\rm Aff}\, (J^1E,\pi^*\Lambda^m\Tan^*M)/\Lambda^m_0\Tan^*E
 \simeq J^1\pi^*$ is more suitable as a dual bundle
 (from the dimensional point of view).
 Then, the canonical forms $\Theta$ and $\Omega$ in ${\cal M}\pi$,
 and $\hat\Theta$ and $\hat\Omega$ in $J^1E^*$,
 can be pulled-back to the restricted and reduced multimomentum bundles
 $J^1\pi^*$ and $\Pi$, using sections of the projections
 $\mu\colon {\cal M}\pi\to J^1\pi^*$ and
 $\delta\colon J^1E^*\to \Pi$, respectively
 \cite{CCI-91}, \cite{EMR-99}.
 In this way, the reduced and restricted multimomentum bundles are
 endowed with (non-canonical) geometrical structures
 ({\sl Hamilton-Cartan forms}\/) needed for
 stating the Hamiltonian formalism.

 In addition, connections in the bundle
 $\pi\colon E\to M$ induce linear sections of $\mu$ (and $\delta$)
 \cite{CCI-91}, \cite{EMR-99}, \cite{GMS-97}, \cite{Sd-95},
 and it can be proved that there is a
 bijective correspondence between the set of
 connections in the bundle $\pi\colon E\to M$, and the set of linear
 sections of the projection $\mu$.

 Hence, all these results, together
 with Theorem \ref{teoisom}, allows us to relate two of the most
 usual Hamiltonian formalisms of Field Theories \cite{EMR-99}.
 \eit

\section{Regular and singular systems}

Following the well-known terminology of mechanics, we define:

\begin{definition}
Let $\ls$ be a Lagrangian system.
\ben
\item
$\ls$ is said to be a {\rm regular} or {\rm non-degenerate}
 Lagrangian system
if ${\cal F}\Lag$, and hence, ${\rm F}\Lag$ are local diffeomorphisms.

As a particular case,
$\ls$ is said to be a {\rm hyper-regular} Lagrangian system
if ${\cal F}\Lag$, and hence ${\rm F}\Lag$, are global diffeomorphisms.
\item
Elsewhere $\ls$ is said to be a {\rm singular} or
 {\rm degenerate} Lagrangian system.
\een
\end{definition}

\begin{prop}
Let $\ls$ a hyper-regular Lagrangian system. Then:
\ben
\item
$\widehat{{\rm F}\Lag}(J^1E)$ is a
$m^2$-codimensional imbedded submanifold of $J^1E^*$
which is transverse to the projection $\delta$.
\item
$\widehat{{\cal F}\Lag}(J^1E)$ and $\widetilde{{\cal F}\Lag}(J^1E)$ are
1-codimensional imbedded submanifolds of ${\cal M}\pi$
which are transverse to the projection $\mu$.
\item
The manifolds
$J^1\pi^*$, $\widehat{{\cal F}\Lag}(J^1E)$,
$\widetilde{{\cal F}\Lag}(J^1E)$,
 $\widehat{{\rm F}\Lag}(J^1E)$ and $\Pi$ are diffeomorphic.

Hence, $\widehat{{\rm F}\Lag}$, $\widehat{{\cal F}\Lag}$
 and $\widetilde{{\cal F}\Lag}$
are diffeomorphisms on their images; and the maps $\mu$,
restricted to $\widehat{{\cal F}\Lag}(J^1E)$ or to
$\widetilde{{\cal F}\Lag}(J^1E)$, and $\iota_0$ and $\delta$,
restricted to $\widehat{{\rm F}\Lag}(J^1E)$, are also
diffeomorphisms. \een \label{hrprop}
\end{prop}
\proof
\ben
\item
If $\Lag$ is hyper-regular then ${\rm F}\Lag$ is a diffeomorphism and
hence,
as ${\rm F}\Lag=\widehat{{\rm F}\Lag}\circ\delta$,
we obtain that $\widehat{{\rm F}\Lag}$ is injective and $\widehat{{\rm
F}\Lag}(J^1E)$
is transverse to the fibers of $\delta$.
\item
The proof of this statement is like for the one above (see also
\cite{LMM-96}).
\item
It is a direct consequence of the above items.
\een
\qed

In this way we have the following (commutative) diagram
\beq
\begin{array}{ccccccc}
\begin{picture}(15,180)(0,0)
\put(0,85){\mbox{$J^1E$}}
\end{picture}
&
\begin{picture}(65,180)(0,0)
\put(20,137){\mbox{${\cal F}\Lag$}}
\put(35,118){\mbox{$\widetilde{{\cal F}\Lag}$}}
\put(36,88){\mbox{$\widehat{{\cal F}\Lag}$}}
\put(36,51){\mbox{$\widehat{{\rm F}\Lag}$}}
\put(20,30){\mbox{${\rm F}\Lag$}}
\put(0,93){\vector(1,1){70}}
\put(0,89){\vector(2,1){70}}
\put(0,85){\vector(1,0){70}}
\put(0,81){\vector(3,-1){70}}
\put(0,77){\vector(1,-1){70}}
\end{picture}
&
\begin{picture}(35,180)(0,0)
\put(16,0){\mbox{$\Pi$}}
 \put(9,42){\mbox{$J^1E^*$}}
\put(12,85){\mbox{${\cal M}\pi$}}
 \put(12,129){\mbox{${\cal M}\pi$}}
 \put(11,170){\mbox{$J^1\pi^*$}}
\put(20,142){\vector(0,1){25}}
 \put(6,98){\vector(0,1){68}}
 \put(20,56){\vector(0,1){25}}
\put(20,36){\vector(0,-1){25}}
 \put(11,152){\mbox{$\mu$}}
\put(-2,140){\mbox{$\mu$}}
\put(11,64){\mbox{$\iota_0$}}
 \put(11,20){\mbox{$\delta$}}
\end{picture}
&
\begin{picture}(10,180)(0,0)
\put(0,165){\vector(0,-1){155}}
 \put(5,85){\mbox{${\mit\Psi}$}}
\end{picture}
\end{array}
\label{diag1} \eeq
 Observe that there exists a map
 $\mu'\colon \widehat{{\cal F}\Lag}(J^1E)\subset{\cal M}\pi
 \to \widetilde{{\cal F}\Lag}(J^1E)\subset{\cal M}\pi$,
 which is a diffeomorphism defined by the relation
 $\widetilde{{\cal F}\Lag}=\mu'\circ\widehat{{\cal F}\Lag}$,
 and $\mu\circ\mu'=\mu$ on $\widehat{{\cal F}\Lag}(J^1E)$.

For dealing with singular Lagrangians we must assume minimal
``regularity'' conditions.
Hence we introduce the following terminology:

\begin{definition}
A singular Lagrangian system $\ls$ is said to be {\rm almost-regular}
if:
\ben
\item
 ${\cal P}:={\cal F}\Lag (J^1E)$ and $P:={\rm F}\Lag (J^1E)$ are closed
submanifolds
 of $J^1\pi^*$ and $\Pi$, respectively.

 (We will denote by $\jmath_0\colon {\cal P}\hookrightarrow J^1\pi^*$
 and  $\j_0\colon P\hookrightarrow \Pi$ the corresponding imbeddings).
\item
 ${\cal F}\Lag$, and hence ${\rm F}\Lag$, are submersions onto their
 images.
\item
For every $\bar y\in J^1E$, the fibers
${\cal F}\Lag^{-1}({\cal F}\Lag (\bar y))$ and hence ${\rm
F}\Lag^{-1}({\rm F}\Lag (\bar y))$
are connected submanifolds of $J^1E$.
\een
\end{definition}

(This definition is equivalent to that in reference \cite{LMM-96},
but slightly different from that in references \cite{GMS-97} and
\cite{Sd-95}).

Let $\ls$ be an almost-regular Lagrangian system. Denote $$
\hat{\cal P}:=\widehat{{\cal F}\Lag}(J^1E) \quad ,\quad
\tilde{\cal P}:=\widetilde{{\cal F}\Lag}(J^1E) \quad ,\quad \hat
P:=\widehat{{\rm F}\Lag}(J^1E) $$ Let $\hat\jmath_0\colon\hat{\cal
P}\hookrightarrow{\cal M}\pi$, $\tilde\jmath_0\colon\tilde{\cal
P}\hookrightarrow{\cal M}\pi$, $\hat{\j}_0\colon\hat
P\hookrightarrow J^1E^*$ be the canonical inclusions, and $$
\hat\mu\colon\hat{\cal P}\to{\cal P} \quad ,\quad
\tilde\mu\colon\tilde{\cal P}\to{\cal P} \quad ,\quad
\hat\iota_0\colon\hat P\to\hat{\cal P} \quad ,\quad
\hat\delta\colon \hat P\to P \quad ,\quad
{\mit\Psi}_0\colon\hat{\cal P}\to P $$ the restrictions of the
maps $\mu$, $\iota_0$, $\delta$ and the diffeomorphism
${\mit\Psi}$, respectively. Finally, define the restriction
mappings $$ {\cal F}\Lag_0\colon J^1E\to{\cal P} \ ,\
\widetilde{{\cal F}\Lag}_0\colon J^1E\to\tilde{\cal P} \ ,\
\widehat{{\cal F}\Lag}_0\colon J^1E\to\hat{\cal P} \ ,\
\widehat{{\rm F}\Lag}_0\colon J^1E\to\hat P \ ,\ {\rm
F}\Lag_0\colon J^1E\to P $$

\begin{prop}
Let $\ls$ be an almost-regular Lagrangian system. Then:
\ben
\item
 The maps ${\mit\Psi}_0$ and $\tilde\mu$ are diffeomorphisms.
\item
For every $\bar y\in J^1E$,
\beq
\widetilde{{\cal F}\Lag_0}^{-1}(\widetilde{{\cal F}\Lag_0}(\bar y))=
{\cal F}\Lag_0^{-1}({\cal F}\Lag_0(\bar y))=
{\rm F}\Lag_0^{-1}({\rm F}\Lag_0(\bar y))
\label{uno}
\eeq
\item
$\tilde{\cal P}$ and  $\hat{\cal P}$ are submanifolds of ${\cal M}\pi$,
$\hat P$ is a submanifold of $J^1E^*$, and
$\tilde\jmath_0\colon\tilde{\cal P}\hookrightarrow{\cal M}\pi$,
$\hat\jmath_0\colon\hat{\cal P}\hookrightarrow{\cal M}\pi$,
$\hat{\j}_0\colon\hat P\hookrightarrow J^1E^*$ are imbeddings.
\item
The restriction mappings
$\widetilde{{\cal F}\Lag}_0$, $\widehat{{\cal F}\Lag}_0$ and
$\widehat{{\rm F}\Lag}_0$
are submersions with connected fibers.
\een
\label{arprop}
\end{prop}
\proof ${\mit\Psi}_0$ is a diffeomorphism as is ${\mit\Psi}$.

 The second equality of (\ref{uno}) is a consequence of the relation
 ${\rm F}\Lag_0={\mit\Psi}_0\circ{\cal F}\Lag_0$.

For the proof of the first equality of (\ref{uno}), and of the
assertions concerning $\tilde\mu$, $\tilde{\cal P}$  and
$\widetilde{{\cal F}\Lag}_0$, see \cite{LMM-96} and
\cite{LMM-96b}. Then, the proofs of the other assertions are
similar. \qed

Thus we have the (commutative) diagram
\beq
\begin{array}{ccccccc}
\begin{picture}(15,180)(0,0)
\put(0,85){\mbox{$J^1E$}}
\end{picture}
&
 \begin{picture}(65,180)(0,0)
 \put(20,137){\mbox{${\cal F}\Lag_0$}}
 \put(35,118){\mbox{$\widetilde{{\cal F}\Lag_0}$}}
 \put(36,91){\mbox{$\widehat{{\cal F}\Lag_0}$}}
 \put(36,51){\mbox{$\widehat{{\rm F}\Lag_0}$}}
 \put(20,30){\mbox{${\rm F}\Lag_0$}}
 \put(0,93){\vector(1,1){70}}
 \put(0,89){\vector(2,1){70}}
 \put(0,87){\vector(1,0){70}}
 \put(0,81){\vector(3,-1){70}}
 \put(0,77){\vector(1,-1){70}}
\end{picture}
&
\begin{picture}(25,180)(0,0)
\put(17,0){\mbox{$P$}} \put(17,42){\mbox{$\hat P$}}
\put(17,85){\mbox{$\hat{\cal P}$}} \put(17,129){\mbox{$\tilde{\cal
P}$}} \put(17,170){\mbox{${\cal P}$}}
\put(20,142){\vector(0,1){25}} \put(20,98){\vector(0,1){25}}
\put(5,98){\vector(0,1){68}} \put(20,56){\vector(0,1){25}}
\put(20,36){\vector(0,-1){25}} \put(11,152){\mbox{$\tilde\mu$}}
\put(-5,140){\mbox{$\hat\mu$}} \put(7,109){\mbox{$\hat\mu'$}}
\put(11,64){\mbox{$\hat\iota_0$}} \put(11,20){\mbox{$\hat\delta$}}
\end{picture}
&
\begin{picture}(6,180)(0,0)
\put(0,165){\vector(0,-1){155}}
\put(3,65){\mbox{${\mit\Psi}_0$}}
\end{picture}
&
 \begin{picture}(50,180)(0,0)
 \put(0,171){\vector(1,0){50}}
 \put(22,175){\mbox{$\jmath_0$}}
 \put(0,131){\vector(1,0){50}}
 \put(22,135){\mbox{$\tilde\jmath_0$}}
 \put(0,90){\vector(1,0){50}}
 \put(22,94){\mbox{$\hat\jmath_0$}}
 \put(0,46){\vector(1,0){50}}
 \put(22,50){\mbox{$\hat{\j}_0$}}
 \put(0,2){\vector(1,0){50}}
 \put(22,6){\mbox{$\j_0$}}
\end{picture}
&
\begin{picture}(35,180)(0,0)
\put(16,0){\mbox{$\Pi$}}
 \put(9,42){\mbox{$J^1E^*$}}
\put(12,85){\mbox{${\cal M}\pi$}}
 \put(12,129){\mbox{${\cal M}\pi$}}
 \put(11,170){\mbox{$J^1\pi^*$}}
\put(20,142){\vector(0,1){25}}
 \put(6,98){\vector(0,1){68}}
 \put(20,56){\vector(0,1){25}}
\put(20,36){\vector(0,-1){25}}
 \put(11,152){\mbox{$\mu$}}
\put(-2,140){\mbox{$\mu$}}
\put(11,64){\mbox{$\iota_0$}}
 \put(11,20){\mbox{$\delta$}}
\end{picture}
&
\begin{picture}(10,180)(0,0)
\put(0,165){\vector(0,-1){155}}
 \put(5,85){\mbox{${\mit\Psi}$}}
\end{picture}
\end{array}
\label{diag2}
\eeq
where $\hat\mu'\colon \hat{\cal P}\to \tilde{\cal P}$
 is defined by the relation
$\hat\mu':=\tilde\mu^{-1}\circ\hat\mu$.

{\bf Remarks}:
 \bit
\item
The fact that $\tilde\mu$ is a diffeomorphism is particularly
relevant, since it allows us to construct a Hamiltonian formalism
for an almost-regular Lagrangian system \cite{LMM-96b}.
\item
It is interesting to point out that the map $\hat\mu$ (which is
related with the Legendre map $\widehat{{\cal F}\Lag}_0$) is not a
diffeomorphism in general, since ${\rm rank}\, \widehat{{\cal
F}\Lag}_0\geq {\rm rank}\, \widetilde{{\cal F}\Lag}_0= {\rm
rank}\, {\cal F}\Lag_0$, as is evident from the analysis of the
corresponding Jacobian matrices.
 \eit

The matrix of the tangent maps ${\cal F}\Lag_*$ and ${\rm
F}\Lag_*$ in a natural coordinate system is $$ \left(\matrix{{\rm
Id} & 0 & 0 \cr 0 & {\rm Id} & 0 \cr
\frac{\partial^2\lag}{\partial x^\nu\partial v^A_\mu} &
\frac{\partial^2\lag}{\partial y^B\partial v^A_\mu} &
\frac{\partial^2\lag}{\partial v^B_\nu\partial v^A_\mu}
\cr}\right) $$ where the sub-matrix
\dst\left(\frac{\partial^2\lag}{\partial v^B_\nu\partial
 v^A_\mu}\right)\) is the {\sl partial Hessian matrix} of $\Lag$.
 Obviously, the regularity of $\Lag$ is equivalent to demanding
 that the partial Hessian matrix
\dst\left(\frac{\partial^2\lag}{\partial v^B_\nu\partial
 v^A_\mu}\right)\)
 is regular everywhere in $J^1E$.
 This fact establishes the relation to the concept of regularity
 given in an equivalent way by saying that
 a Lagrangian system $\ls$ is {\sl regular} if
 $\Omega_{\Lag}$ is $1$-nondegenerate (elsewhere it is said to be
 {\sl singular} or {\sl non-regular}).

\section*{Conclusions}

\begin{itemize}
  \item
  We have reviewed the definitions of four different
  multimomentum bundles for the Hamiltonian formalism of first-order
  Classical Field Theories (multisymplectic models).
  The so-called {\sl generalized} and {\sl reduced} multimomentum
bundles
  are related straightforward from their definition,
  and the same thing happens with the
  {\sl generalized} and {\sl restricted} multimomentum bundles.
  The first goal of this work has been to relate both couples,
  proving that the {\sl reduced} and {\sl restricted} multimomentum
bundles
  are, in fact, canonically diffeomorphic. In natural local
  coordinates, this diffeomorphism is just the identity.
  \item
  The canonical forms which the generalized and the extended
  multimomentum bundles are endowed with, have been defined and
  characterized in several equivalent ways.
  \item
    Given a {\sl Lagrangian system} in Field Theory,
  we have introduced the corresponding {\sl Legendre maps}
  relating these multimomentum bundles to the first-order jet
  bundle associated with this system. Some of them,
  the {\sl generalized} and {\sl reduced} Legendre maps,
  are defined in a natural way as {\sl fiber derivatives}
  of the Lagrangian density, being the other ones
  obtained from those.
  The relation among all these maps has been clarified.
  \item
  {\sl Regular} and {\sl almost-regular} Lagrangian systems are
  defined and studied, attending to the geometric
  features of the Legendre maps.
  In this way, the standard definitions existing in the usual
  literature are extended and completed.
\end{itemize}

\appendix

\section{Appendix}

\begin{lem}
Let $\pi\colon F\to N$ be a differentiable bundle, with $\dim\,
N=n$ and $\dim\, F=n+r$, and $p\in F$ with $q=\pi(p)$. Let
$\moment{v}{1}{h}\in\Tan_pF$ ($h\leq n$), such that:
$\moment{v}{1}{h}$ are linearly independent, and
$\langle\moment{v}{1}{h}\rangle\cap{\rm V}_p(\pi)=\{ 0\}$. Then:
\ben
\item
There exist $\moment{X}{1}{h}\in\vf (W)$, for a neighborhood $W\subset
F$ of $p$,
such that:
\ben
\item
$\moment{X}{1}{h}$ are linearly independent, at every point of $W$.
\item
$\moment{X}{1}{h}$ generate an involutive distribution in $W$.
\item
$X_i(p)=v_i$, ($i=1,\ldots ,h$).
\item
$\langle X_1(x),\ldots ,X_h(x)\rangle\cap{\rm V}_x(\pi)=\{ 0\}$, for
every $x\in W$.
\een
\item
There exists a local section $\gamma$ of $\pi$, defined in a
neighborhood of
$q\in N$, and $\moment{u}{1}{h}\in\Tan_qN$ such that:
$\gamma (q)=p$, and
$\Tan_p\gamma (u_i)=v_i$, ($i=1,\ldots ,h$).
\een
\label{lema0}
\end{lem}
\proof
Let $\moment{v}{h+1}{n}\in\Tan_pF$, such that
$\Tan_p={\rm V}_p(\pi)\oplus\langle\moment{v}{1}{n}\rangle$.
Let $(W,\varphi)$ be a local chart of $F$ at $p$, adapted to $\pi$.
We have
$\varphi\colon W\to U_1\times U_2\subset\Real^n\times\Real^r$.
Let $\moment{e}{1}{n}$ and $\moment{e}{n+1}{n+r}$ be local basis of
$\Real^n$ and $\Real^r$, respectively. Then
\bea
\langle\Tan_p\varphi(v_1),\ldots ,\Tan_p\varphi(v_n)\rangle =
\langle\moment{e}{1}{n}\rangle
\label{ayuda}
\\
\Tan_x\varphi({\rm V}_x(\pi)) =\langle\moment{e}{n+1}{n+r}\rangle
\quad ,\quad \forall x\in W
\nonumber
\eea
Let $\moment{Z}{1}{n}\in\vf (U_1\times U_2)$ be the constant extensions
of
$\Tan_p\varphi(v_1),\ldots ,\Tan_p\varphi(v_n)$ to $U_1\times U_2$. We
have that
$[Z_i,Z_j]=0$, $\forall i,j$. Finally, let
$\moment{X}{1}{n}\in\vf (W)$, with $X_i=\varphi^*Z_i$, therefore:
\ben
\item
Taking $\moment{X}{1}{h}$, ($h\leq n$),
conditions (1.a), (1.b) and (1.c) hold trivially (by construction), and
condition (1.d) holds as a consequence of (\ref{ayuda}).
\item
Observe that $\moment{X}{1}{n}\in\vf (W)$ generate the horizontal
subspace of a connection defined in the bundle $\pi\colon
W\to\pi(W)$, which is integrable because the distribution is
involutive. Then, let $\gamma$ be the integral section of this
connection at $p$. Therefore $\gamma(q)=p$, and the subspace
tangent to the image of $\gamma$ at $p$ is generated by
$\moment{v}{1}{n}$. Hence, there exist $\moment{u}{1}{n}$ such
that $\Tan_q\gamma(u_i)=v_i$. \een \qed

Now, let $A$ be an affine space modeled on a vector space $S$, and $T$
another vector space, both over the same field $K$.
Let ${\rm Aff}(A,T)$ be the set of affine maps from $A$ to $T$; that is,

maps $\varphi\colon A\to T$ such that there exists
a linear map $\hat\varphi\colon S\to T$ verifying that
$\varphi (a)-\varphi (b)=\hat\varphi (a-b)$; for $a,b\in A$.
Then:

\begin{lem}
\ben
\item
There is a natural isomorphism between
${\rm Aff}(A,T)/T$ and $S^*\otimes T$
(and then
$\dim\,{\rm Aff}(A,T)=\dim\, (S^*\otimes T)+\dim\, T=\dim\, T(\dim\,
S+1)$).
\item
There is a canonical isomorphism between
${\rm Aff}(A,T)$ and ${\rm Aff}(A,K)\otimes T$.
\een
\label{isocan}
\end{lem}
\proof ${\rm Aff}(A,T)$ is a vector space over $K$ with the
natural operations. The map $\wedge \colon{\rm Aff}(A,T)\to
S^*\otimes T$, which assigns $\hat\varphi$ to every $\varphi$, is
linear and we have the exact sequence $$ 0\longrightarrow T
\stackrel{j}{\longrightarrow} {\rm
Aff}(A,T)\stackrel{\wedge}{\longrightarrow}S^*\otimes T
\longrightarrow 0 $$ where, if $t\in T$, then $j(t)\colon A\to T$
is the constant map $(j(t))(a)=t$, for every $a\in A$. Therefore
we have a natural isomorphism ${\rm Aff}(A,T)/T\simeq S^*\otimes
T$ and then $$ \dim\,{\rm Aff}(A,T)=\dim\, (S^*\otimes T)+\dim\,
T=\dim\, T(\dim\, S+1) $$ Moreover, for $y_0\in A$, there exists
an splitting ${\rm Aff}\, (A,T)\simeq T\oplus (S^*\otimes T)$
given by the following retract of the above exact sequence $$
\begin{array}{cccccc}
j_{y_0}&\colon& S^*\otimes T &\to &{\rm Aff}\, (A,T) &
\\
& & \varphi & \mapsto & \varphi_0\colon & y\mapsto \varphi (y-y_0)
\end{array}
$$
On the other hand, we have the bilinear map $$
\begin{array}{cccc}
{\rm Aff}(A,K)\times T & \longrightarrow & {\rm Aff}(A,T) &
\\
(\alpha ,t) & \mapsto & t\alpha \ \colon & a \mapsto \alpha (a)t
\end{array}
$$
and hence we can define the following morphism
\beann
{\rm Aff}(A,K)\otimes T & \longrightarrow & {\rm Aff}(A,T)
\\
\alpha^i\otimes u_i & \mapsto & u_i\alpha^i \eeann which is
injective because we can assume that the vectors $u_i$ are
linearly independent, and both spaces have the same dimension.
Therefore ${\rm Aff}(A,T)$ and ${\rm Aff}(A,K)\otimes T$ are
canonically isomorphic. \qed

If $(\moment{s}{1}{m})$ is a basis of $S$,
$(\coor{\sigma}{1}{m})$ is its dual basis and $(\moment{t}{1}{m})$
is a basis of $T$, then taking an affine reference in $A$,
$(t_i,t_i\otimes\sigma^j)$ is a basis of ${\rm Aff}(A,T)$, as vector
space.

Now, let $G,H$ be finite dimensional vector spaces over $K$,
and $F$ a subspace of $G$. Consider the exact sequence
\beq
0\longrightarrow F \stackrel{\tau}{\longrightarrow}G
\stackrel{\pi}{\longrightarrow}H\longrightarrow 0
\label{sequence}
\eeq
The set
$$
\Sigma\equiv\{ \sigma\colon H\to G\ ;\ \sigma\ {\rm linear}\ ,
\pi\circ\sigma ={\rm Id}_H\}
$$
is an affine space modeled on $L(H,F)=H^*\otimes F$.
In fact, if $\sigma\in \Sigma$ and $\lambda\in H^*\otimes F$,
then $\sigma +\lambda\in\Sigma$, since $\pi\circ\lambda =0$.
Furthermore, if $\sigma,\mu\in\Sigma$, then $\pi\circ (\sigma -\mu)=0$,
and hence $\sigma-\mu\in H^*\otimes F$.

If $\dim\, H=m$, taking the space ${\rm Aff}(\Sigma ,\Lambda^mH^*)$, and

according to the second item of the above Lemma, we have that
${\rm Aff}(\Sigma ,\Lambda^mH^*)\simeq
{\rm Aff}(\Sigma ,K)\otimes\Lambda^mH^*$,
and then
$\dim\, {\rm Aff}(\Sigma ,\Lambda^mH^*)=\dim\,{\rm Aff}(\Sigma ,K)=
\dim\, (H^*\otimes F)+1$.
Now, consider the subspace
$$
\Lambda_1^mG^*\equiv
\{\alpha\in\Lambda^mG^*\ :\ \inn(u)\inn(v)\alpha =0\ ,\ u,v\in F\}
\subset\Lambda^mG^*
$$

\begin{lem}
The spaces ${\rm Aff}(\Sigma,\Lambda^mH^*)$ and $\Lambda_1^mG^*$
are canonically isomorphic.
\label{isocan1}
\end{lem}
\proof
If $(\coor{f}{1}{r})$ is a basis of $F$ and
$\coor{g}{1}{m}\in G$ such that $(\coor{f}{1}{r},\coor{g}{1}{m})$
is a basis of $G$, then $\Lambda_1^mG^*$ is generated by
$(g^1\wedge\ldots\wedge g^m,f^j\wedge g^{i_1}\wedge\ldots\wedge
g^{i_{m-1}})$
(with $1\leq i_1<\ldots <i_{m-1}<m$),
therefore $\dim\, \Lambda_1^mG^*=1+\dim\, F\,\dim\, H$;
that is, $\dim\, \Lambda_1^mG^*=\dim\,{\rm Aff}(\Sigma,\Lambda^mH^*)$.
Next, define the linear map
$$
\begin{array}{ccccccc}
\Upsilon&\colon&\Lambda_1^mG^*&\longrightarrow&{\rm
Aff}(\Sigma,\Lambda^mH^*)& &
\\
& & \eta & \mapsto & \Upsilon (\eta) & \colon & \sigma \mapsto
\sigma^*\eta
\end{array}
$$ which we want to prove is an isomorphism, for which it suffices
to prove that it is injective. Thus, suppose that $\Upsilon
(\eta)=0$ (that is $\sigma^*\eta=0$, for every $\sigma\in\Sigma$),
then we must prove that $\eta=0$. Let $\coor{g}{1}{m}\in G$ be
linearly independent (so $\dim\, L(\coor{g}{1}{m})=m$); we are
going to calculate $\eta (\coor{g}{1}{m})$. \ben
\item
If $L(\coor{g}{1}{m})\cap \tau  (F)=\{ 0\}$:

Then $G=L(\coor{g}{1}{m})\oplus \tau  (F)$ and therefore
$\pi\colon  L(\coor{g}{1}{m})\to H$ is an isomorphism.
Hence, there exist $\coor{h}{1}{m}\in H$ and
$\sigma\colon H\to G$, with $\pi\circ\sigma={\rm Id}_H$,
such that $\sigma (h^i)=g^i$; therefore
$$
\eta (\coor{g}{1}{m})=
\eta (\sigma(h^1),\ldots ,\sigma(h^m))=
(\sigma^*\eta) (\coor{h}{1}{m})=0
$$
\item
If $L(\coor{g}{1}{m})\cap \tau  (F)\not=\{ 0\}$:
\ben
\item
If $\dim\, (L(\coor{g}{1}{m})\cap \tau  (F))=1$:

In this case there is $u\in L(\coor{g}{1}{m})\cap \tau  (F)$, with
$u\not= 0$,
such that $L(\coor{g}{1}{m})=L(u,\coor{g}{2}{m})$
(up to an ordering change),
and then there is $k\in K$ such that
$$
\eta (\coor{g}{1}{m})=k\eta (u,\coor{g}{2}{m})
$$
therefore $L(\coor{g}{2}{m})\cap \tau  (F)=\{ 0\}$.
Let $(u,\coor{u}{1}{r})$ be a basis of $F$, and $\bar g^1$ such that
$(u,\coor{u}{1}{r},\bar g^1,\coor{g}{2}{m})$ is a basis of $G$.
Then $L(\bar g^1,\coor{g}{2}{m})\cap \tau  (F)=\{ 0\}$,
and hence, as a consequence of the above item, we conclude that
$\eta (\bar g^1,\coor{g}{2}{m})=0$.
Furthermore, $L(\bar g^1+u,\coor{g}{2}{m})\cap \tau  (F)=\{ 0\}$
because, if this is not true, then
$L(\bar g^1,\coor{g}{2}{m})\cap \tau  (F)\not=\{ 0\}$, and hence
$\eta(\bar g^1+u,\coor{g}{2}{m})=0$, by the above item. Thus
$$
\eta(\coor{g}{1}{m})=k\eta(u,\coor{g}{2}{m})=
k(\eta(\bar g^1+u,\coor{g}{2}{m})-\eta(\bar g^1,\coor{g}{2}{m}))=0
$$
\item
If $\dim\, (L(\coor{g}{1}{m})\cap \tau  (F))=s>1$, with $s\leq r=\dim\,
F$:

Let $(\coor{f}{1}{s},\coor{g}{s+1}{m})$ be a basis of
$L(\coor{g}{1}{m})$, with $\coor{f}{1}{s}\in F$. Then, as
$\eta\in\Lambda^mG^*$,
$$
\eta(\coor{g}{1}{m})=k\eta(\coor{f}{1}{s},\coor{g}{s+1}{m})=0
$$
\een
\een
Then $\eta =0$, so $\Upsilon$ is injective and is a canonical
isomorphism between
${\rm Aff}(\Sigma,\Lambda^mH^*)$ and $\Lambda_1^mG^*$.
\qed

Using the first item of Lemma \ref{isocan}
and identifying $A=\Sigma$, we conclude:

\begin{lem}
${\rm Aff}(\Sigma,\Lambda^mH^*)/\Lambda^mH^*\simeq
(H^*\otimes F)^*\otimes\Lambda^mH^*\simeq
H\otimes F^*\otimes\Lambda^mH^*$.
\label{isocan2}
\end{lem}

\subsection*{Acknowledgments}

We are grateful for the financial support of the CICYT
TAP97-0969-C03-01. We wish to thank Mr. Jeff Palmer for his
assistance in preparing the English version of the manuscript.

\end{document}